\documentclass[prb,aps,showkeys,twocolumn,showpacs,preprintnumbers,amsmath,amssymb]{revtex4}

\renewcommand{\vec}{\bm}
\renewcommand{\phi}{\varphi}

\usepackage{graphicx}
\usepackage{dcolumn}
\usepackage{bm}

\begin{document}


\title{Anisotropic spin splitting and spin relaxation in asymmetric zinc-blende semiconductor quantum structures}

\author{J.~Kainz}
\email{josef.kainz@physik.uni-regensburg.de}
\author{U.~R\"ossler}%
\affiliation{
Institut f\"ur Theoretische Physik, Universit\"at Regensburg,
93040 Regensburg, Germany
}%

\author{R.~Winkler}
\affiliation{
Institut f\"ur Technische Physik III, Staudtstr. 7,
D-91058 Erlangen, Germany
}%

\date{\today}

\begin{abstract}
  Spin relaxation due to the D'yakonov-Perel' mechanism is
  intimately related with the spin splitting of the electronic
  states. We determine the spin relaxation rates from anisotropic
  spin splittings of electron subbands in n-(001) zinc-blende
  semiconductor quantum structures calculated self-consistently in the
  multi-band envelope function approach.  The giant anisotropy of
  spin relaxation rates found for different spin-components in the
  (001) plane can be ascribed to the interplay between the bulk and
  quantum well inversion asymmetry.  One of the in-plane relaxation
  rates may exhibit a striking nonmonotonous dependence on the
  carrier density.
\end{abstract}

\maketitle

\section{Introduction}

The spin degree of freedom of electrons in solids has recently
attracted much interest
\cite{divincenzo:1995,prinz:1998,kikkawa:1998,kikkawa:1999,kikkawa:1997}
in the perspective of spintronic devices, which require long spin
relaxation times. Thus, a quantitative understanding of the
dependence of spin-relaxation times on the system parameters is
needed for the desired engineering of systems with the required
properties as well as for the interpretation of measured values.
\cite{tackeuchi:1996,terauchi:1999,malinowski:2000,adachi:2001} It
is generally accepted \cite{pikus:1984} that in bulk zinc-blende
type semiconductors and in quantum well (QW) structures based on these 
materials electronic spin relaxation is governed by the
D'yakonov-Perel' \cite{dyakonov:1971,dyakonov:1972} (DP) and the
Bir-Aronov-Pikus \cite{bir:1976} mechanisms. The
former mechanism becomes dominant in n-doped (001)-grown QW systems,
which we consider here. 
The DP mechanism is intimately related with the
spin splitting of the electronic subbands caused by the lacking 
inversion symmetry. \cite{averkiev2:2002}
It can result from the bulk crystal
structure (bulk inversion asymmetry, BIA \cite{dresselhaus:1955}) 
but also from the QW structure 
(structural inversion asymmetry, SIA \cite{bychkov:1984}).
Making use of this relation, Averkiev et al.~\cite{averkiev:1999,averkiev2:2002}
employ a simple variational calculation for a triangular 
potential model of n-doped GaAs heterostructures and include 
BIA and SIA terms as a perturbation to calculate spin-splitting 
and DP spin-relaxation rates. The simplicity of this approach raises
the question of the reliability of these results in view of a more 
rigorous calculation, as will be presented in this paper.

In this contribution, we determine the electron spin
relaxation rates for longitudinal (in growth direction) 
and transverse (in-plane) spin components in 
asymmetric n-doped (001)-grown zinc-blende semiconductor quantum
structures using the 
anisotropic spin splitting obtained from self-consistent calculations
in the multi-band envelope-function approximation (EFA).
\cite{winkler:1993}
We investigate the dependence of these rates on the QW width 
and carrier concentration for the material systems 
AlGaAs/GaAs and AlGaSb/InAs, and compare our results for the former system 
with those of Ref.~\onlinecite{averkiev2:2002}.

\section{DP mechanism and spin splitting}

The DP spin relaxation is described in a single band approach by
considering the time evolution of the electron spin-density vector
\begin{equation}
\vec{S}=\langle \mathrm{tr}(\rho_{\vec{k}} \vec{\sigma}) \rangle=
\sum_{\vec{k}} \mathrm{tr}(\rho_{\vec{k}} \vec{\sigma}),   
\end{equation}
where
$\vec{k}$ is the electron wave vector, $\vec{\sigma}$ the vector of
Pauli spin matrices, and $\rho_{\vec{k}}$ the $2 \times 2$ electron
spin-density matrix.
The dynamics of $\rho_{\vec{k}}$ is ruled by a Bloch equation
containing the spin-orbit coupling $H_\mathrm{so}=\hbar \vec{\sigma}
\cdot \vec{\Omega}(\vec{k}) $ and the (spin independent) momentum
scattering, e.g., with phonons, (nonmagnetic) impurities or other
electrons \cite{Glazov:2002,brand:2002}
characterized by a
momentum scattering time $\tau_p$. The spin-orbit coupling is
analogous to a Zeeman term, but with an effective magnetic field
$\vec{\Omega}(\vec{k})$ that depends on the underlying material and
geometry and on the electron wave vector $\vec{k}$.  The vector
$\mathrm{tr}(\rho_{\vec{k}} \vec{\sigma}) $ precesses about
$\vec{\Omega}(\vec{k})$ which eventually results in spin relaxation.
However, momentum scattering changes $\vec{k}$ and thus the
direction and magnitude of $\vec{\Omega}(\vec{k})$ felt by the
electrons.  Therefore, frequent
scattering (on the time scale of $| \vec{\Omega}(\vec{k}) |^{-1}$)
suppresses the precession and consequently the spin relaxation. This
is the motional-narrowing behavior that is typical for the DP
mechanism, according to which the spin relaxation rate $\tau_s^{-1}
\propto \tau_p$. \cite{dyakonov:1972}

Due to time reversal invariance we have $\vec{\Omega}(0) = 0$. At
finite $\vec{k}$ the spin-orbit coupling $H_\mathrm{so}$ causes a
spin splitting $2 \hbar |\vec{\Omega}(\vec{k})|$ of the subband
structure. It exists already in bulk semiconductors with broken
inversion symmetry. In this article we focus on semiconductors with
a zinc-blende structure where the BIA spin splitting of the electron
states is characterized in leading order by the Dresselhaus or $k^3$ term
\cite{dresselhaus:1955,braun:1985}
\begin{equation}
\vec{\Omega}_\mathrm{BIA}(\vec{k})= \frac{\gamma}{\hbar}
\left(\begin{array}{c}
k_x \left(k_y^2 - k_z^2\right) \\[0.5ex]
k_y \left(k_z^2 - k_x^2\right) \\[0.5ex]
k_z \left(k_x^2 - k_y^2\right)
  \end{array} \right)
\label{omega_BIA}
\end{equation}
where $\gamma$ is a material dependent parameter.
In the following we will assume
that the $z$ axis is perpendicular to the QW plane. In asymmetric
QW's SIA gives rise to a second contribution to spin-orbit coupling. Its leading
term, linear in $\vec{k}_\parallel = ( k_x, k_y, 0 ) $ is frequently
called the Rashba term. \cite{bychkov:1984} It is characterized by
the effective field
\begin{equation}
\vec{\Omega}_\mathrm{SIA}(\vec{k}_\parallel)= 
\frac{\alpha}{\hbar} \left(\begin{array}{c}
    k_y \\[0.5ex]
    -k_x \\[0.5ex]
    0
  \end{array} \right)
\label{omega_SIA}
\end{equation}
with a prefactor $\alpha$ that depends on the material parameters of the
underlying semiconductor bulk material (like $\gamma$) but in
addition also on the asymmetry of the QW in growth
direction.
In first order perturbation theory for (001) oriented QW's the terms $k_z$ (and powers thereof) in the BIA
term are replaced by expectation values with respect to the subband function,
thus one has \cite{dyakonov:1986}
\begin{equation}
\label{omega_formel}
\vec{\Omega}(\vec{k}_\parallel)
=
\frac{\gamma}{\hbar} \left(\begin{array}{c}
    k_x \left(k_y^2 - \langle k_z^2 \rangle\right) \\[0.5ex]
    k_y \left(\langle k_z^2 \rangle - k_x^2\right) \\[0.5ex]
    0 
    \end{array} \right) +
\vec{\Omega}_\mathrm{SIA}(\vec{k}_\parallel)
\, .
\end{equation}

Within the DP mechanism the relaxation of the components of
$\vec{S}$ are ruled by \cite{dyakonov:1972}
\begin{equation}
\frac{\partial}{\partial t} S_i = - \frac{1}{\tau_i} S_i
\end{equation}  
with $i=z,+,-$ corresponding to the components of $\vec{S}$ along
$[001]$, $[110]$, and $[\overline{1}10]$, $S_\pm = 1/\sqrt{2}
( S_x \pm S_y )$, and spin-relaxation rates
\cite{averkiev2:2002}
\begin{subequations}
\label{tau_formel}
\begin{eqnarray}
\frac{1}{\tau_z} & = & \frac{4 \tau_p}{\hbar^2} 
 \biggl[ \left( \gamma^2 \langle k_z^2 \rangle^2 + \alpha^2 \right) k_F^2 
 - \frac{\gamma^2 \langle k_z^2 \rangle k_F^4}{2}
 \nonumber \\ && 
  + \frac{\gamma^2 k_F^6}{8}  \biggr] 
 \;\;\; \label{tau_z_formel}  \\[1ex]
%
\frac{1}{\tau_\pm} & = & \frac{2 \tau_p}{\hbar^2} 
 \biggl[ k_F^2 \left( \pm \alpha - \gamma \langle k_z^2 \rangle \right) 
 \left(
 \pm \alpha - \gamma \langle k_z^2 \rangle + \frac{\gamma}{2} k_F^2
   \right) 
\nonumber
\\ && +
 \frac{ \gamma^2 k_F^6}{8} \biggr]
 \; .
 \label{tau_minus_formel}  
\end{eqnarray}  
 \label{tau_pm_formel}  
\end{subequations}
Here we have assumed an isotropic momentum scattering and a
degenerate electron system. The rates (\ref{tau_pm_formel}) depend
explicitly on the
carrier density $n$ via the radius $k_F$ of the
Fermi circle and implicitly via the self-consistent subband functions in the
expectation values of $k_z^2$.
\begin{figure}
\includegraphics[width= 0.41 \textwidth]{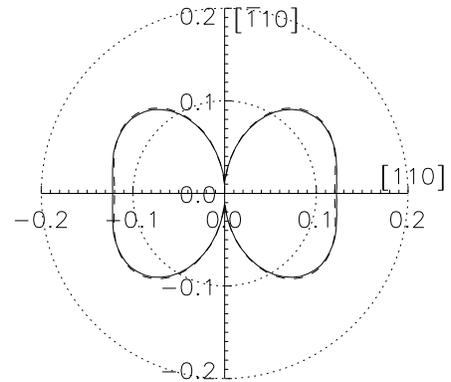}
\caption{\label{spinsplit_fig} Polar plot of the calculated (full
line) and fitted (dashed line) spin splitting $2 \hbar
|\vec{\Omega}(\vec{k}_\parallel)|$ in $\mathrm{meV}$ for $\vec{k}_\parallel$ at the
Fermi contour in an asymmetrically doped (001) AlAs/GaAs
heterostructure with charge density $n=3 \times 10^{11}
\mathrm{cm}^{-2}$.}
\end{figure}

In a previous evaluation of Eq.~(\ref{tau_pm_formel}) in 
n-doped (001) grown GaAs heterostructures, Averkiev et al.~\cite{averkiev2:2002}
employed the triangular potential model and the variational subband function
of Ref.~\onlinecite{fang:1966}, thus assuming an infinite barrier at the interface
 and working in the single-band approximation.
In their model, the electric field varies linearly with the carrier
density $n$, which results in $\alpha \propto n$. Using Eq.\ 
(\ref{tau_formel}) and a tabulated value of $\gamma =
27\,\mathrm{eV}$\AA\ (appropriate for bulk GaAs) these authors
calculate the relaxation rates for carrier densities up to $2.5
\times 10^{13} \, \mathrm{cm}^{-2}$. (Their results for a smaller
range of $n$ are reproduced in Fig.~\ref{tau_golub_fig}.) However,
carrier densities in real AlAs/GaAs quantum structures with
occupation of the lowest subband only are limited to $n \lesssim 8
\times 10^{11} \, \mathrm{cm}^{-2}$ (see Ref.\ 
\onlinecite{hamilton:1995} and discussion below). Moreover, the
approach of Averkiev et al. does not yield full self-consistency nor
does it include nonparabolic corrections or effects due to the
barrier material, which have to be taken into account for finite
barriers.

\begin{figure}
\includegraphics[width= 0.40 \textwidth]{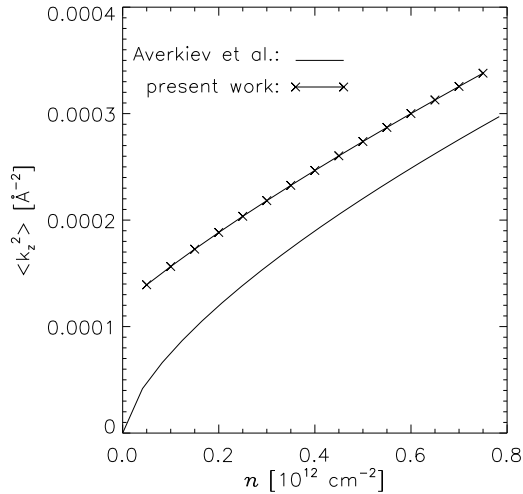}
\caption{\label{kappaquadrat_fig} 
Comparison of data from Ref. \onlinecite{averkiev2:2002} with present results for the 
expectation value $\langle k_z^2 \rangle$ as a function of charge density $n$ for
an asymmetrically doped AlAs/GaAs heterostructure.
}
\end{figure}
\begin{figure}
\includegraphics[width= 0.40 \textwidth]{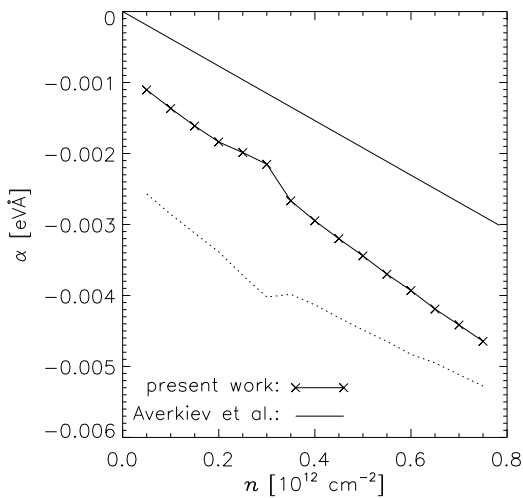}
\caption{\label{rashba_fig} 
Comparison of data from Ref.\
\onlinecite{averkiev2:2002} with present results for the 
Rashba coefficient $\alpha$ as a function of charge density $n$ for
an asymmetrically doped AlAs/GaAs heterostructure. For comparison of
SIA and $k$ linear BIA terms the dotted line shows the $k$
linear part of the BIA term $- \gamma \langle k_z^2 \rangle$.
}
\end{figure}
\begin{figure}
\includegraphics[width= 0.40 \textwidth]{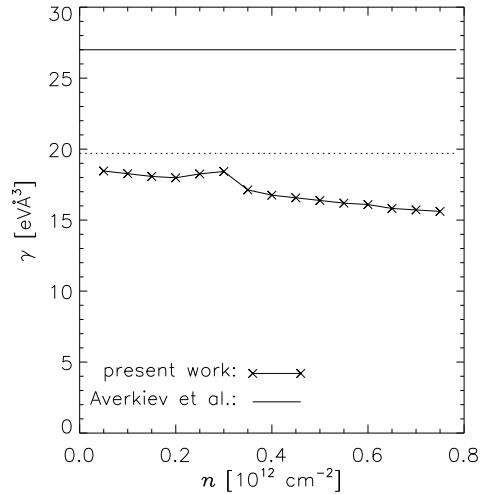}
\caption{\label{gamma_fig} 
Comparison of data from Ref. \onlinecite{averkiev2:2002} with present results for the 
BIA coefficient $\gamma$ as a function of charge density $n$ for
an asymmetrically doped AlAs/GaAs heterostructure. The dotted line indicates the value for
bulk GaAs.
}
\end{figure}
In order to overcome these deficiencies we perform fully
self-consistent calculations of the subband spin splitting in the
multiband EFA (considering five levels or 14 bands
\cite{winkler:1993}) in application to (001) oriented AlGaAs/GaAs
and AlGaSb/InAs QW's, n-doped on one side.  Such calculations
include in a systematic way the BIA nonparabolicity modified by the
QW confinement and the SIA or Rashba term. They have successfully
explained the anisotropic spin splitting measured with inelastic
light-scattering. \cite{wissinger:1998} Figure~\ref{spinsplit_fig}
shows a typical polar plot of the calculated spin splitting in the
$(k_x,k_y)$ plane (solid contour). The striking feature is the
pronounced anisotropy with the strongly reduced spin splitting in
the $[\overline{1}10]$ direction.
From the numerically calculated spin splitting
$\vec{\Omega}_\mathrm{num} (\vec{k}_\parallel)$ we extract the
parameters $\alpha$ and $\gamma$ by fitting
$\vec{\Omega}_\mathrm{num} (\vec{k}_F)$ to Eq.\ (\ref{omega_formel})
using $k_F= \sqrt{2 \pi n}$. The expectation value $\langle k_z^2
\rangle$ is evaluated using the EFA wave functions. The quality of
this approach is demonstrated by the dashed contour in
Fig.~\ref{spinsplit_fig}. In Fig.~\ref{kappaquadrat_fig} we show
$\langle k_z^2 \rangle$ as a function of the density $n$ together
with the estimate $\langle k_z^2 \rangle \propto n^{2/3}$ used in
Ref.~\onlinecite{averkiev2:2002}. Figures~\ref{rashba_fig} and
\ref{gamma_fig} show how the parameters $\alpha$ and $\gamma$ change
as a function of $n$.\footnote{The kinks in Figs.~\ref{rashba_fig}
and \ref{gamma_fig} around $n \approx 0.3 \times
10^{12}\mathrm{cm}^{-2}$ are due to higher order terms that
are included in the EFA calculations, but not accounted for in the
model of Eq.\ (\ref{omega_formel}).} For $\alpha$ this dependence is
expected since $\alpha$ contains the expectation value of the
electric field.  Our more realistic calculations yield absolute
values of the Rashba coefficient that are substantially larger than
those used in Ref.~\onlinecite{averkiev2:2002}.
The dependence of $\gamma$ on $n$ (which is only moderate) 
reflects
higher order nonparabolicity corrections with increasing Fermi
energy known from the bulk material. \cite{roessler:1984}
For our set of band parameters, the bulk values of $\gamma$ are
$19.7 \, \mathrm{eV\AA^3}$ for GaAs\footnote{The difference between the
value of $\gamma=27\,\mathrm{eV}$\AA\ from Ref.~\onlinecite{averkiev2:2002}
and our value is due to taking into account the coupling parameter $\Delta^-$ 
(see Ref.~\onlinecite{mayer:1991}).
}
and $18.5 \, \mathrm{eV\AA^3}$ for AlAs.

Recently Lau et al.\cite{Lau:2001} have performed similar
calculations, yet for symmetrically doped QW's in order to compare with
experimental data obtained from multiple QW's. \cite{terauchi:1999}
Our focus here is on asymmetric single QW's and the interplay between SIA
and BIA terms, which require fully self-consistent EFA calculations.

The simpler 8-band model, which takes into account BIA only
perturbatively, yields qualitatively the same spin splitting as the
14-band Hamiltonian. However, the magnitude of the spin splitting is
about 10--20\% larger in the 8-band model. \cite{wissinger:1998}
Therefore, all results given in this paper are based on the more
realistic 14-band model.

\begin{figure}
\includegraphics[width= 0.40 \textwidth]{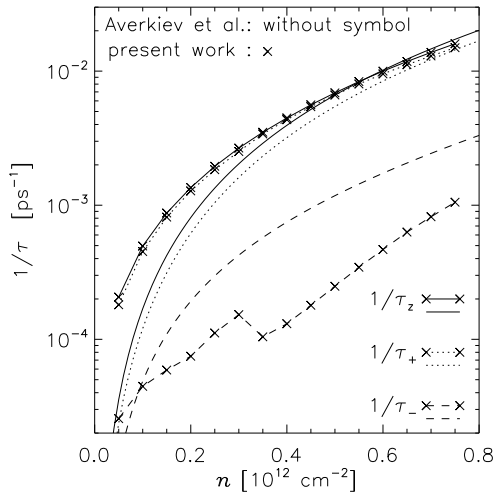}
\caption{\label{tau_golub_fig} Comparison of data from
Ref.~\onlinecite{averkiev2:2002} with present results (marked ``$\times$'')
for the spin relaxation rates $\tau_z^{-1}$, $\tau_+^{-1}$,
$\tau_-^{-1}$ as a function of charge density $n$ for an
asymmetrically doped AlAs/GaAs heterostructure. The range of $n$ is
chosen so that only the lowest (spin-split) subband is occupied.}
\end{figure}

\section{Calculated spin relaxation rates}
\begin{figure*}
\includegraphics[width= \textwidth]{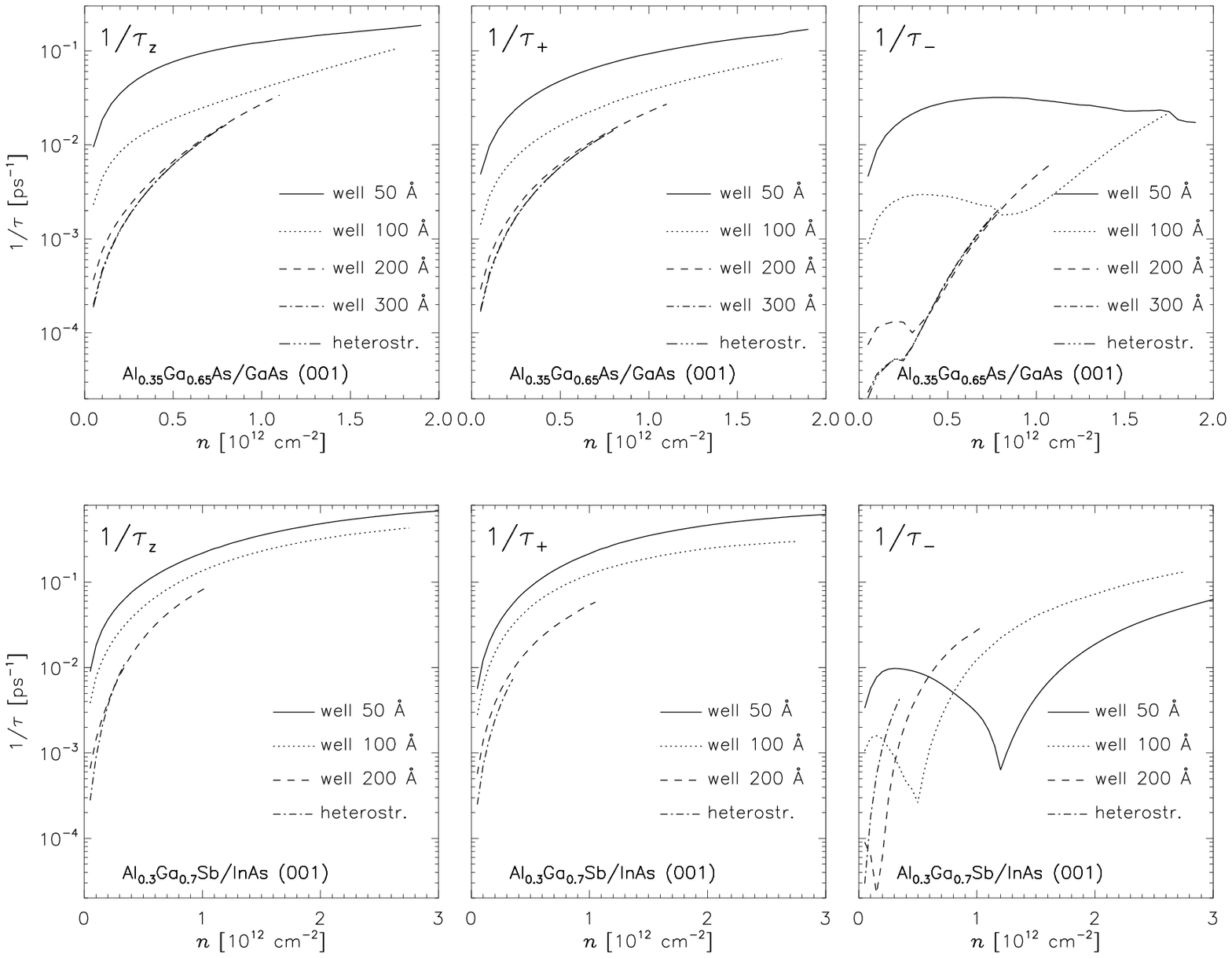}
\vspace*{-8ex}
\caption{\label{tau_fig} Spin relaxation rates $\tau_z^{-1}$,
$\tau_+^{-1}$, $\tau_-^{-1}$ as a function of charge density $n$ for
Al$_{0.35}$Ga$_{0.65}$As/GaAs (001) (upper row) and
Al$_{0.3}$Ga$_{0.7}$Sb/InAs (001) (lower row) quantum wells with
widths $L=$ 50, 100, 200$\,$\AA ~and for a simple heterostructure.
The range of $n$ is chosen so that only the lowest (spin split)
subband is occupied.}
\end{figure*}
In this section we present our results for the spin-relaxation rates
obtained from Eq.\ (\ref{tau_formel}) with the parameters 
$\langle k_z^2 \rangle$, $\alpha$ and
$\gamma$ determined within the 14-band
EFA calculations. This is done by assuming a momentum-scattering
time $\tau_p = 0.1 \, \mathrm{ps}$. As all rates are proportional to
$\tau_p$ they may be readily rescaled. For all calculations
presented in this paper the range of $n$ is chosen such that only
the lowest (spin-split) subband is occupied. In
Fig.~\ref{tau_golub_fig} our results are shown for a AlAs/GaAs
heterostructure in comparison with the data from Ref.\ 
\onlinecite{averkiev2:2002}. (Note that in Ref.\ 
\onlinecite{averkiev2:2002} $\tau_+$ and $\tau_-$ are interchanged
\footnote{
We choose our coordinate system such that the electric field
points in $[001]$ direction, whereas in Ref.~\onlinecite{averkiev2:2002}
the electric field points in the opposite direction, which leads to 
an interchange of $\tau_+$ and $\tau_-$.
}.)  
Both in Ref.~\onlinecite{averkiev2:2002} and in our calculation,
the rates $\tau_z^{-1}$ and
$\tau_+^{-1}$ increase monotonously with the carrier density and are
close to each other over the whole range.
For the higher densities ($n \gtrsim 0.4 \times
10^{12} \mathrm{cm}^{-2}$) our results for $\tau_z^{-1}$ and
$\tau_+^{-1}$ coincide with the respective data from Ref.\ 
\onlinecite{averkiev2:2002}, whereas our relaxation rates are
generally much larger for small $n$. This can be attributed to the
fact that in Ref.\ \onlinecite{averkiev2:2002} it was assumed that
$\langle k_z^2 \rangle \propto n^{2/3}$, while in our calculations
$\langle k_z^2 \rangle$ does not tend towards zero for 
small $n$ (Fig.~\ref{kappaquadrat_fig}).
Our results for $\tau_-^{-1}$ are generally much smaller than those
of Ref.\ \onlinecite{averkiev2:2002} and show also a pronounced
nonmonotonous behavior around $n \approx 0.4 \times 10^{12}
\mathrm{cm}^{-2}$, which is not present\footnote{
The nonmonotonous dependence visible in Fig.~3 of Ref.~\onlinecite{averkiev2:2002}
is a consequence of the approximation that a Boltzmann distribution
is used for an almost degenerate electron system.}
in Ref.\ \onlinecite{averkiev2:2002}. 
A detailed explanation of the origin of
this feature will be given below. The small oscillations below and
above  the dip at $n \approx 0.4 \times 10^{12}
\mathrm{cm}^{-2}$ are numerical noise.

Spin relaxation rates calculated as a function of the QW width $L$
and carrier density $n$ are shown in Fig.~\ref{tau_fig} for 
single-side n-doped QW's with (001) orientation for the material
systems Al$_{0.35}$Ga$_{0.65}$As/GaAs and
Al$_{0.3}$Ga$_{0.7}$Sb/InAs. 
With increasing $L$ the asymmetric single-side doped
QW's become more and more similar to a single heterostructure, and therefore
the results for $L \gtrsim 200\,$\AA~are almost the same as for a
heterostructure. Similar to the GaAs/AlAs system of
Fig.~\ref{tau_golub_fig} the rates $\tau_z^{-1}$ and $\tau_+^{-1}$
are close to each other in all cases, while $\tau_-^{-1}$ is
distinctly smaller (frequently more than an order of magnitude) and
shows the nonmonotonous dependence on $n$.  
Both the GaAs and the InAs QW's
show qualitatively the same behavior.


\begin{figure}
\includegraphics[width= 0.40 \textwidth]{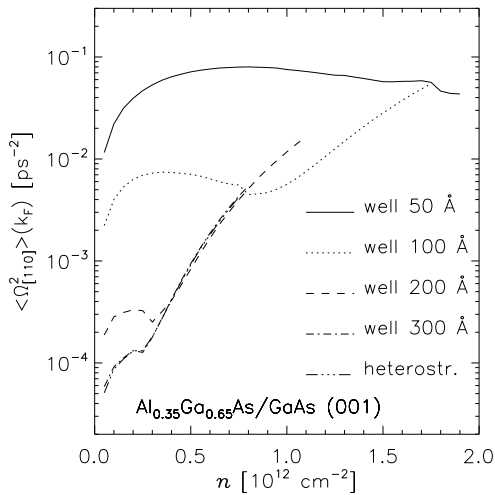}
\caption{\label{transverse_omega_fig} 
Mean squared transverse field 
$\langle \Omega^2_{[110]}\rangle (k_F)$ as a function of $n$ for the
AlGaAs/GaAs systems of Fig.~\ref{tau_fig}. }
\end{figure}

In Fig.~\ref{tau_fig} we have a dip in the density dependence of
$\tau_-^{-1}$ that is most pronounced for the InAs QW's.
This feature, not seen in Ref.~\onlinecite{averkiev2:2002},
appears here due to the differences in $\langle k_z^2 \rangle$,
$\alpha$ and $\gamma$ (see Figs. \ref{kappaquadrat_fig},\ref{rashba_fig},\ref{gamma_fig}).
They cause in our calculations the first term in the expression
defining $\tau_-^{-1}$ [Eq. (\ref{tau_minus_formel})] to become negative
in a certain range of $n$ and therefore lead to a nonmonotonous behavior.

The origin of the dip in $\tau_-^{-1}$ can be understood more clearly
within a picture of the DP mechanism, where we assume
that the spin relaxation rate for the spin component in the direction $\vec{e}$ 
is only effected by those components of the effective
field $\vec{\Omega}(\vec{k}_\parallel)$ that are perpendicular to $\vec{e}$.
In analogy with the behavior of spin relaxation rates for nuclear magnetic resonance
\cite{slichter:1963} the spin relaxation rate $\tau_-^{-1}$ of the spin component 
in the direction $[\overline{1}10]$ should be proportional to the average of the squared 
effective field components perpendicular to that direction:
\begin{equation}
 \tau_-^{-1} \propto \left( \langle \Omega_{[110]}^2 \rangle (k_F) + 
 \langle \Omega_{[001]}^2 \rangle (k_F) \right) 
 \; ,
\end{equation}
where $\langle \Omega_{[110]}^2 \rangle (k_F)$ represents an average
over the Fermi contour
\begin{subequations}
\begin{equation}
\langle \Omega_{[110]}^2 \rangle (k_F) =
 \frac{1}{2 \pi} \int_0^{2 \pi} d\phi \: \Omega_{[110]}^2 (\vec{k}_F (\phi)) 
\end{equation}
and $\Omega_{[110]} (\vec{k}_F (\phi))$ denotes a projection of
$\vec{\Omega} (\vec{k}_F (\phi))$ on the direction $[110]$
\begin{eqnarray}
\Omega_{[110]} (\vec{k}_F (\phi)) & = & \vec{\Omega} (\vec{k}_F (\phi)) \cdot 
\left(\begin{array}{c}
1/\sqrt{2} \\
1/\sqrt{2} \\
0
\end{array} \right)
\; , \\
\vec{k}_F(\phi) & = & k_F 
\left(\begin{array}{c}
\cos{\phi} \\
\sin{\phi} \\
 0
\end{array} \right) \; .
\end{eqnarray}  
\end{subequations}
As the projection of $\vec{\Omega}(\vec{k}_\parallel)$ onto the $[001]$ direction vanishes, 
$\langle \Omega_{[001]}^2 \rangle (k_F)$ is equal to zero. Hence, $\langle \Omega_{[110]}^2 \rangle (k_F)$
is a measure of the relaxation rate $\tau_-^{-1}$. 
In fact, by calculating
this quantity as a function of $n$ (Fig.~\ref{transverse_omega_fig})
we obtain the same behavior as for $\tau_-^{-1}$,
thus confirming the given interpretation. 
In the same line of arguments $\tau_+^{-1}$ is essentially determined by
$\langle \Omega^2_{[\overline{1}10]}\rangle (k_F)$ and, in fact, this
quantity exhibits a monotonous dependence on $n$ similar to $\tau_+^{-1}$
in Figs.~\ref{tau_golub_fig} and \ref{tau_fig}.

\section{Proposed Measurements}
Our calculations indicate a large difference of the spin relaxation times for
spins oriented in $\pm$ and $z$ direction. This feature is quite universal; it appears both
in GaAs- and InAs-based systems over a wide range of well widths and carrier densities.
In a certain density range (''dips'' in Fig.~\ref{tau_fig}) the
spin relaxation times can differ by more than two orders of magnitude.
Therefore, we expect that this difference can be observed experimentally.

The creation of spins oriented in $z$ direction using optical orientation techniques and the
measurement of their relaxation has already been 
demonstrated. \cite{tackeuchi:1996,terauchi:1999,malinowski:2000,adachi:2001}
For the creation of an in-plane spin orientation a spin-injection technique
similar to the paradigmatic spin transistor \cite{datta:1990} can be used.
Here the carrier density can be varied by means of a gate 
electrode and the spin relaxation can be detected via the source-drain current.

Another way to determine in-plane spin relaxation is provided by 
measurements of the Hanle effect \cite{averkiev:2002}.

We propose a second scheme for the creation of in-plane excess spins and 
the measurement of their relaxation time: In a first step, spins oriented 
in $z$ direction are created with conventional optical orientation
techniques. These can be rotated into the plane of the QW by means of a so-called
tipping pulse, as demonstrated by Gupta \emph{et al.} \cite{gupta:2001}
After a certain delay, the spins are rotated back into the $z$
direction, so that the magnitude
of the remaining spin polarization can be measured. By changing the
delay time the spin relaxation for an arbitrary in-plane direction can be analyzed.
A substantial difference between the spin relaxation times can be expected
for spins oriented along the in-plane directions [110] and $[\overline{1}10]$.

We propose that
such a system can be used as a ``spin storage'': The rotation of
spins from the $z$ direction into the $[\overline{1}10]$ direction effectively
stores the spins because the relaxation time for the latter  direction is larger by orders
of magnitude.

\section{Conclusion}
Starting from a realistic model for the anisotropic spin splitting in
GaAs/AlGaAs and InAs/AlGaSb quantum structures we calculate the
parameters which determine the longitudinal ($\tau_z^{-1}$) and
transverse ($\tau_\pm^{-1}$) spin-relaxation rates and determine
their dependence on QW width and carrier concentration. We quantify
the previously predicted \cite{averkiev:1999} giant anisotropy of
spin relaxation by our more realistic calculation of
the spin splitting. In contrast with previous investigations
our accurate calculations predict a pronounced nonmonotonous
behavior of $\tau_-^{-1}(n)$, which can be understood by ascribing the
spin relaxation to the transverse components of the effective magnetic field.
Future work will include calculations for different
material systems, growth directions and doping profiles. For
comparison with room-temperature experiments the temperature
dependence and the effects of different scattering mechanisms have
to be considered. Furthermore, the microscopic interface asymmetry,
that has been identified as an additional mechanism of spin
splitting, \cite{roessler:2002} will be taken into account.

\begin{acknowledgments}
We would like to thank L.~E.~Golub for valuable discussions.
This work has been supported by the DFG via Forschergruppe~370 {\em
Ferromagnet-Halbleiter-Nanostrukturen}.
\end{acknowledgments}

\bibliography{mybib}

\end{document}